\documentclass[lettersize,journal]{IEEEtran}

\usepackage{amsmath,amsfonts}
\usepackage{graphicx}
\usepackage{cite}
\usepackage{array}
\usepackage{url}
\usepackage{xurl}

\usepackage[caption=false,font=normalsize,labelfont=sf,textfont=sf]{subfig}
\usepackage{pifont}
\usepackage{algpseudocode}
\usepackage{multirow} 
\usepackage{booktabs}
\begin{document}

\title{FlashAccel: Leveraging High-Bandwidth Flash (HBF) for High-Throughput LLM Inference}

\author{Xinyu~Wang, Yalong~Xue, Xiaotian~Sun, Xiaoyu~Zhang, Xinjiang~Zhang, Chunmeng~Dou,~\IEEEmembership{Member,~IEEE}, Xueqi~Li,~\IEEEmembership{Member,~IEEE}, and~Xiaoming~Chen,~\IEEEmembership{Member,~IEEE}
\thanks{This work was supported in part by the National Natural Science Foundation of China under Grant 62488101 and Grant 62495104, and in part by the Youth Innovation Promotion Association CAS.  (\textit{Corresponding author: X. Chen})
}
\thanks{
X. Wang is with the Institute of Computing Technology, Chinese Academy of Sciences, Beijing 100190, China, and also with the University of Chinese Academy of Sciences, Beijing 101408, China (email: wangxinyu22s@ict.ac.cn).
}
\thanks{
Y. Xue is with the School of Advanced Interdisciplinary Sciences, University of Chinese Academy of Sciences, Beijing 101408, China, and also with the Institute of Computing Technology, Chinese Academy of Sciences, Beijing 100190, China (email: xueyalong22@mails.ucas.ac.cn).
}
\thanks{
C. Dou is with the Institute of Microelectronics, Chinese Academy of Sciences, Beijing 100029, China (e-mail: douchunmeng@ime.ac.cn).
}
\thanks{
X. Sun, Xiaoyu Zhang, Xinjiang Zhang, X. Li, and X. Chen are with the Institute of Computing Technology, Chinese Academy of Sciences, Beijing 100190, China (email: sunxiaotian21s@ict.ac.cn; zhangxiaoyu@ict.ac.cn; zhangxinjiang@ict.ac.cn; lixueqi@ict.ac.cn; chenxiaoming@ict.ac.cn).
}

}

\maketitle

\begin{abstract}
Large language model (LLM) inference is increasingly limited by the capacity of High-Bandwidth Memory (HBM) in GPUs, as model weights and KV cache grow rapidly. High-Bandwidth Flash (HBF) provides higher capacity than HBM while offering comparable bandwidth, making it a promising substrate for capacity-constrained LLM inference. However, its inherently high access latency, low bandwidth utilization, and lack of support for heterogeneous resource management make it difficult to integrate HBF into GPUs for LLM inference. We present FlashAccel, a co-designed system that enables efficient LLM inference using HBF. FlashAccel integrates HBF into HBM-based GPUs, providing architectural support to mitigate access latency. It improves bandwidth utilization through specialized data layouts for both model weights and KV cache, and introduces an HBF-aware storage management layer together with a programming model to organize persistent data in HBF and coordinate heterogeneous memory resources at the system level. 
Experimental results demonstrate that integrating six HBF stacks into the GPU enables FlashAccel to deliver an average improvement of 2.49$\times$ and 1.93$\times$ in throughput per GPU and energy efficiency over the HBM-only GPU under a 100ms latency constraint, respectively.

\end{abstract}

\begin{IEEEkeywords}
High-Bandwidth Flash, LLM Inference, Heterogeneous Memory.
\end{IEEEkeywords}

\section{Introduction}

\begin{figure}[t]
    \centering
    \includegraphics[width=0.78\linewidth]{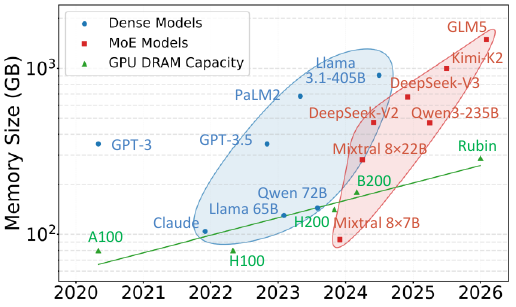}
    \caption{Trends in HBM memory capacity and model size.}
    \label{Fig::Size}
\end{figure}

The rapid advancement of large language models (LLMs) is driving inference workloads toward multi-turn interactions~\cite{deshpande2025multichallenge}, agents~\cite{yao2022react}, and long-context processing~\cite{bai2024longbench}. Larger models expand weight footprint, while longer contexts increase KV cache size~\cite{bhaskar2025cache}.
These trends fundamentally increase the memory capacity requirements of inference systems. However, the capacity of High-Bandwidth Memory (HBM) in GPUs has not kept pace with the rapid growth of memory demand. Figure~\ref{Fig::Size} shows that, although GPU memory capacity has increased steadily over time, the memory footprint of popular models has grown much faster. Today, models can exceed terabytes in size, far surpassing the capacity of a single GPU, which is at most a few hundred gigabytes.
Meanwhile, context lengths have increased by 4$\times$ on average over the past two years~\cite{aubakirova2026state}, proportionally increasing KV cache storage requirements.

The limited HBM capacity introduces several critical challenges in LLM inference. First, during the memory-bound decode phase, insufficient memory capacity restricts the batch size, leading to low compute utilization and throughput. Second, the memory pressure often forces early eviction of KV cache, preventing reuse opportunities in multi-turn interactions and increasing recomputation overhead~\cite{kwon2023vllm}. Third, due to the limited capacity of a single GPU, storing large models requires multi-GPU systems, which incur significant hardware cost and introduce additional complexity in inter-GPU communication, task scheduling, and fault tolerance.

Prior works~\cite{yu2024cambricon,sun2025lincoln,lee2025aif} have leveraged Flash memory to address the capacity bottleneck. They typically store the model weights in Flash and use in-storage computation to exploit internal parallelism. 
However, these works are primarily designed for edge scenarios. Their limited bandwidth and computational capacity are insufficient to meet the requirements of high-throughput scenarios that require powerful compute devices (i.e., GPUs) and memory devices with both large capacity and high bandwidth.
To provide both high bandwidth and large capacity, High-Bandwidth Flash (HBF) has been proposed. HBF is a specialized Flash memory that delivers bandwidth comparable to HBM while retaining the high density and non-volatility of Flash~\cite{ha2026h,hsu2026haven}.
By integrating HBF into the GPU alongside HBM, we can significantly expand memory capacity to achieve potentially higher decode throughput. However, this capacity advantage does not automatically translate into throughput gains.

Several critical challenges prevent leveraging HBF for high-throughput inference.
First, Flash suffers from inherently \textbf{high access latency}~\cite{boukhobza2025survey}, which can increase the end-to-end latency.
Second, although HBF provides high peak bandwidth, it requires a high degree of parallelism to reach the peak bandwidth. Inefficient data layout can limit parallel access opportunities and result in \textbf{low bandwidth utilization}. Third, integrating HBF introduces heterogeneous memory resources, raising the \textbf{challenge of managing these resources and exposing efficient abstractions to software}.

Addressing these challenges requires hardware–software co-design. We propose FlashAccel, an accelerator system that co-designs architecture, data layout, and system software to efficiently integrate HBF into GPU-based LLM inference.
Architecturally, FlashAccel mitigates Flash access latency with SRAM cache distributed across HBF dies for data prefetching. It integrates HBF, HBM, and GPU cores into a unified accelerator to relieve the capacity bottleneck.
At the data layout level, FlashAccel addresses the low bandwidth utilization challenge of HBF by designing specialized layouts for weights and KV cache, enabling high parallelism during inference.
At the system level, FlashAccel provides an HBF-aware storage management layer to organize persistent data in HBF and a programming model that allows applications to efficiently utilize heterogeneous memory resources.

In summary, we make the following contributions.
\begin{itemize}
    \item We identify the key challenges of employing HBF for LLM inference, including high access latency that affects the end-to-end latency, the gap between peak read bandwidth and effective bandwidth under LLM workloads, and the lack of management for heterogeneous memory resources.
    \item We introduce FlashAccel, an HBF-based accelerator for high-throughput LLM serving. FlashAccel combines hardware–software co-design techniques that hide high access latency, enable near-peak bandwidth for weights and KV cache, and efficiently manage heterogeneous resources.
    \item Experimental results show that FlashAccel significantly improves throughput and energy efficiency by 2.49$\times$ and 1.93$\times$, respectively, compared with HBM-only GPU implementations.
\end{itemize}

\section{Background \& Motivation}

\subsection{LLM Workload}

LLMs consist of multiple stacked transformer layers, each requiring storage for static model weights (shared across all requests) and dynamic KV cache (isolated per request). 
LLM inference consists of prefill and decode phases with distinct computational characteristics~\cite{kwon2023vllm}. 
Prefill processes all input tokens together, making it predominantly compute-bound. In decode, each request provides one token per inference step, leading to frequent access to the model weights and KV cache, which makes it memory-bound.

As KV cache grows over inference steps, most frameworks~\cite{zheng2024sglang, kwon2023vllm} adopt PagedAttention-style mechanisms~\cite{kwon2023vllm} that partition KV cache into fixed-size blocks for fine-grained memory allocation, reducing fragmentation and improving memory utilization. Frameworks also manage KV cache eviction. Upon request completion, the computed KV cache is retained in memory for potential reuse by future requests. When memory is exhausted, the framework evicts historical KV cache to serve new requests.

\subsection{Memory Capacity Bottleneck}

\begin{figure}[t]
    \centering
    \includegraphics[width=.9\linewidth]{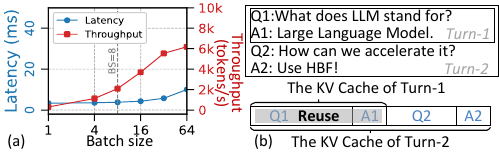}
    \caption{(a) Benefits of batching. All results are obtained using an 8-layer configuration of Qwen3-32B due to capacity limits (For the full model, H100 only supports batch size less than 8). (b) Benefits of KV cache reuse.} 
    \label{Fig::Reuse}
\end{figure}

The rapid growth of model weights and KV cache has made GPUs' HBM capacity a fundamental bottleneck for LLM inference.
This bottleneck directly limits the batch size in the decode phase.
Consider running Qwen3-32B~\cite{yang2025qwen3} on an H100 GPU~\cite{nvidia_h100_datasheet} with 80GB HBM. After reserving approximately 64GB for model weights, only 16GB remains available for KV cache. For an 8K context length, each request requires about 2GB KV cache, meaning the GPU can support a batch size of at most 8. According to the roofline model~\cite{williams2009roofline}, larger batch sizes increase the arithmetic intensity of weight-related operations and thus improve GPU utilization. Figure~\ref{Fig::Reuse}(a) shows that a small batch size limits throughput, whereas larger batch sizes significantly improve throughput with only marginal impact on latency.

HBM capacity further limits the reuse opportunities of KV cache. Figure~\ref{Fig::Reuse}(b) illustrates the reuse opportunity in multi-turn interactions. KV cache generated in previous turns can be reused during the prefill phase for future turns, avoiding recomputation of this KV cache. However, reuse requires retaining KV caches in memory. In the example above, each request requires 2GB of KV cache, meaning a single GPU can store KV caches for only eight requests. When serving more requests, KV caches are frequently evicted, leading to low cache hit rates and additional recomputation overhead. Prior work~\cite{wang2025kvcache} shows that achieving high reuse efficiency may require up to 4$\times$ more memory capacity for KV cache.

Current LLM inference systems mitigate memory capacity limitations by scaling out to multiple GPUs, but this increases deployment cost, system complexity, and inter-GPU communication overhead. Therefore, increasing per-GPU memory capacity while preserving high bandwidth provides a more promising alternative.

\subsection{Flash for LLM Inference}

\begin{figure}[t]
    \centering
    \includegraphics[width=1.0\linewidth]{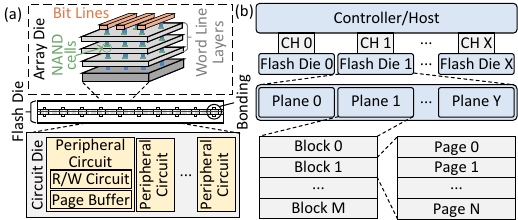}
    \caption{Architecture of Flash memory. (a) Physical structure. (b) Logical structure.}
    \label{Fig::Flash}
\end{figure}

Flash memory is attractive for alleviating the capacity bottleneck due to its high storage density. As shown in Figure~\ref{Fig::Flash}(a), a Flash die comprises an array die stacked over a circuit die via hybrid bonding~\cite{zhou2024research}. The array die stores data using vertically stacked NAND cells, while the circuit die integrates peripheral circuits.
Figure~\ref{Fig::Flash}(b) shows the Flash hierarchy of channel-die-plane-block-page. 
A Flash device consists of multiple channels, each of which serves as an independent interface and is shared by one or more dies. Each die contains multiple planes with independent read/write paths in the circuit die, enabling parallel data exchange between the NAND array and the page buffer.
Within a plane, blocks are the units of erase operation, while pages are the units of read and program operations. 

Recent works such as AiF~\cite{lee2025aif}, Lincoln~\cite{sun2025lincoln}, and Cambricon-LLM~\cite{yu2024cambricon} use Flash for LLM inference. They store model weights in Flash and place compute units within Flash to exploit plane-level parallelism. These designs enable edge inference, but are not suitable for high-throughput inference. These designs provide only hundreds of GB/s bandwidth and few TFLOPS of computational capacity, far below modern GPUs. These limitations in both bandwidth and compute capacity directly restrict inference throughput. Moreover, they store only model weights in Flash but ignore the large KV cache footprint. 
A more promising approach to accelerating LLM inference with Flash is to integrate a Flash device of high bandwidth with the GPU, alleviating its capacity bottleneck. The key to achieving high bandwidth is exploiting  plane-level parallelism. An HBF can be realized by increasing the number of planes per die (e.g., partitioning a die into 16 planes~\cite{sandisk2025memorycentricai}), adopting HBM-like 3D stacking~\cite{jun2017hbm} to enable parallelism across multiple Flash dies, and leveraging TSVs to provide high-bandwidth interconnects between Flash and GPUs.

The main concern with storing KV cache in Flash is endurance. For example, single-level cell (SLC) Flash typically provides about 100K P/E cycles~\cite{mohan2010learned}. 
Prior works~\cite{sun2025lincoln,lee2025aif,yu2024cambricon} avoid storing frequently updated KV caches in Flash to circumvent endurance limits.
However, recent advances in attention mechanisms make storing KV cache in Flash feasible.
According to the DeepSeek report~\cite{deepseek2025v3r1inference}, the prefill cluster (approximately 764 GPUs) writes 266 billion new tokens per day\footnote{The prefill cluster processes 608 billion tokens in total, but 342 billion tokens are served via KV cache hits and do not require additional writes.}. Building upon these inputs, the decode cluster (approximately 1,051 GPUs) generates an additional 168 billion tokens.
Based on these statistics, each GPU in the prefill cluster writes 22.7TB of KV cache per day, while each GPU in the decode cluster writes 26.9TB of KV cache per day.
SLC Flash can provide about 55 P/E cycles per day over a 5-year lifetime. 
Thus, a 1TB HBF device can sustain about 55TB of data writes per day, which is sufficient for the inference workloads.
In addition, relaxing retention guarantees can extend endurance by up to 50$\times$~\cite{luo2015warm,cai2012flash}, making Flash a practical medium for KV cache storage.

\section{Challenges}

\subsection{High Access Latency}
Compared with HBM, Flash exhibits significantly higher access latency. Even with high read bandwidth comparable to HBM, the high read latency of HBF severely degrades end-to-end performance.
Using HBF, the GPU receives the first batch of data roughly 4 $\mu s$ (tR) after issuing an access request~\cite{kouchi2020128gb}, whereas HBM requires only about 100 $ns$~\cite{huang2021shuhai}, which is a 40$\times$ gap.
For example, a 4k $\times$ 4k FP8 GEMV operation can be completed on H200 in approximately 4 $\mu s$. When using HBF instead, it incurs an additional 4 $\mu s$ access latency, leading to a 2$\times$ end-to-end latency increase.

\begin{figure}[t]
    \centering
    \includegraphics[width=1.0\linewidth]{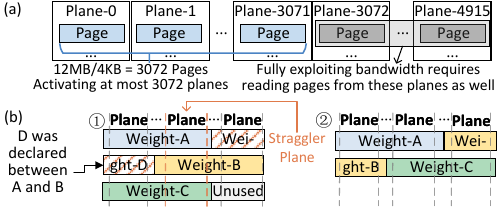}
    \caption{Low bandwidth utilization of accessing weights in HBF. (a) Accessing one weight. (b) Accessing multiple weights.}
    \label{Fig::Weight}
\end{figure}

\begin{figure*}[t]
    \centering
    \includegraphics[width=\textwidth]{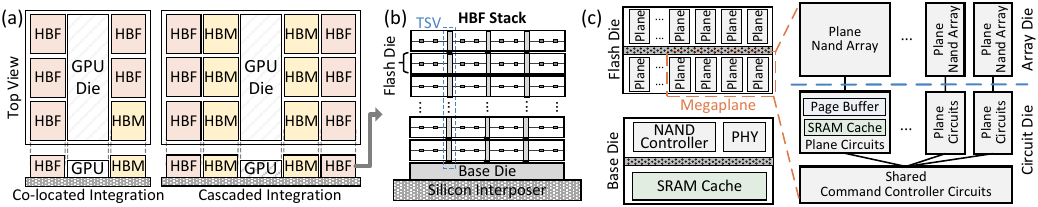}
    \caption{Architecture of FlashAccel. (a) Two options for integrating HBF. (b) Architecture of an HBF stack. (c) Design of array die, circuit die, and base die inside HBF.}
    \label{Fig::Arch}
\end{figure*}

\subsection{Low Bandwidth Utilization}
Achieving high bandwidth in HBF requires massive inter-plane parallelism, because the high read latency of each plane limits its standalone bandwidth. For example, matching the 4.8 TB/s bandwidth of H200 would ideally require 4,916 planes reading the 4KB pages concurrently, with each read taking 4 $\mu s$. 
However, data layout and access patterns often limit the plane-level parallelism, preventing HBF from achieving its peak bandwidth.
Consider an HBF with $P$ planes and peak bandwidth of $B_{\text{peak}}$. The requested data spans $P_{\mathrm{active}}$
planes and places $L_i$ pages on plane $i$, where $L_i$ is the
\textit{plane load}.
Since a plane serves page reads serially, the overall access latency is determined by the maximum plane load $L_{\max}=\max_i L_i$. The total data volume amounts to $P_{\text{active}}\bar{L}$ pages, where $\bar{L}$ is the average plane load. The effective bandwidth is thus
\begin{equation}
B_{\text{eff}} = B_{\text{peak}} \times \frac{P_{\text{active}}}{P} \times \frac{\bar{L}}{L_{\max}}.
\label{Eq::Bandwidth}
\end{equation}
Bandwidth utilization degrades under two conditions. First, if $P_{\text{active}} < P$, the inactive planes contribute no bandwidth and the aggregate peak becomes unreachable. Second, if a straggler plane exists ($L_{\max} > \bar{L}$), the straggler must serve multiple reads while the other planes turn idle early.

\begin{figure}[t]
    \centering
    \includegraphics[width=1.0\linewidth]{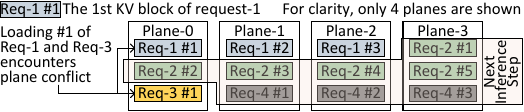}
    \caption{Imbalance of accessing KV cache in HBF.}
    \label{Fig::KVCache}
\end{figure}

Both weight and KV cache accesses suffer from such low utilization. For weights, a single matrix is often too small to span all planes. For example, a 12 MB expert weight matrix in Qwen3-235B occupies and activates only about 3,072 of the 4,916 planes when distributed at page granularity (Figure~\ref{Fig::Weight}(a)). Consecutively accessing multiple weights can activate more
planes. However, the resulting bandwidth still depends on their physical layout.
Existing inference frameworks store weights in source-code declaration order rather than access order. In case \ding{172} of Figure~\ref{Fig::Weight}(b), when weights A, B, and C are accessed consecutively, the declaration order places D between A and B, yielding $L_{\max}=3$. In contrast, reorganizing the weights by execution order (case \ding{173}) reduces $L_{\max}$ to 2 and lowers the access latency by 33\%.

The problem is more severe for the KV cache, whose distribution evolves during inference. At each inference step, the system accesses the KV cache aggregated from all active requests. As requests complete and new ones arrive, an initially balanced distribution cannot be sustained. As shown in Figure~\ref{Fig::KVCache}, requests 1-4 are processed first, with their KV cache evenly distributed across planes\footnote{
Using PagedAttention, the KV cache is split into blocks, and we assume the KV size of a layer in a KV block equals the page size of a plane in the figures.}. In a later step, if requests 1 and 3 complete, only requests 2 and 4 remain active, yielding $L_{\max}=3$ against $\bar{L}=2$, resulting in 33\% bandwidth loss. Over time, the paging and eviction mechanisms further fragment each request's KV cache across multiple planes, while the active request set evolves in a near-random manner, making imbalance the norm. Modeling this distribution as a balls-into-bins problem~\cite{raab1998balls}, distributing the 100GB KV cache of Qwen3-235B across 4,916 planes shows $L_{\max}/\bar{L}=1.52$.
As shown in Figure~\ref{Fig::KVCache}, attention kernels may simultaneously access two requests' KV cache in the same plane, leading to contention analogous to bank conflicts, which further degrades bandwidth utilization.

\subsection{System Resource Management}

The third challenge is system-level support for HBF.
Unlike HBM, HBF is non-volatile, so the system must support persistent data organization inside HBF for objects such as model weights and KV cache. It also has to coordinate HBF, HBM, and SRAM during execution. Existing approaches do not fit the management of HBF well. 
Using an HBM-style paging mechanism for data stored in HBF provides direct and low-latency access, but it does not support persistence, because page table mappings are discarded when the program terminates. 
Conventional Flash stacks support persistence but incur metadata and latency overhead from layered address translation~\cite{agrawal2008design,boukhobza2025survey}.
Moreover, as HBF bandwidth approaches that of HBM, treating Flash as a lower tier behind HBM becomes increasingly inefficient because it forces unnecessary staging through HBM. Instead, HBF should be exposed as a directly accessible resource for GPUs, with SRAM introduced to hide Flash access latency. 
This requires a storage management layer for persistent data in HBF and a programming abstraction for accessing heterogeneous memory resources.

\section{Hardware Design}

FlashAccel integrates HBF, HBM, and GPU cores into a heterogeneous memory accelerator for LLM inference. As shown in Figure~\ref{Fig::Arch}(a), FlashAccel can be implemented using either co-located integration (CLI)~\cite{sandisk2025memorycentricai} or cascaded integration (CSI)~\cite{ha2026h}. CLI places HBF alongside the GPU and replaces part of the HBM, increasing the total memory capacity at a lower cost. CSI connects HBF to the GPU through the HBM base die in a daisy chain manner, preserving more HBM capacity and bandwidth. HBM and HBF serve complementary roles. HBM handles small, frequently updated intermediate data with low latency, while HBF stores large, read-heavy data such as model weights and KV cache. 
Figure~\ref{Fig::Arch}(b) shows the HBF stack architecture. Each stack consists of eight Flash dies and a base die connected through TSVs. Each Flash die integrates an array die and a circuit die using hybrid bonding. Following \cite{sun2025lincoln} and \cite{sandisk2025memorycentricai}, FlashAccel scales each Flash die to 96 planes to increase plane-level parallelism. To amortize the complexity of control logic, these planes are further organized into multiple \textit{megaplanes}, where planes within each megaplane share the same NAND command control circuits. Upon receiving a NAND command, a megaplane simultaneously accesses the data located at identical block and page offsets across all of its constituent planes. Such coupled accesses effectively aggregate the pages or blocks at the same offset in these planes into a larger \textit{megapage} or \textit{megablock} visible to the outside, thereby realizing fine-grained plane-level parallelism. The megaplane becomes the new independent access unit in our design.
For systems with multiple HBF stacks and Flash dies, FlashAccel further introduces hyperpage abstraction. A \textit{hyperpage} is a logical access unit that aggregates one megapage from all megaplanes in all HBF stacks. Unlike the superpage in SSD controllers, which serves for resource allocation and write operations, the hyperpage mainly targets read operations. Accessing HBF at hyperpage granularity enables concurrent accesses across all planes, fully exploiting available plane-level parallelism. 
As shown in Figure~\ref{Fig::Arch}(c), FlashAccel integrates SRAM buffers in both the circuit dies and the base die to hide Flash read latency through prefetching. The unutilized area of each circuit die is used to provide a private SRAM cache for each plane. The base die integrates the NAND controller, PHY, and an additional stack-level SRAM cache. FlashAccel uses SLC-based Flash for its lower read latency and higher write endurance. Although SLC has lower bit density than MLC and TLC Flash, HBF still provides a substantial capacity advantage over HBM.

\section{Data Layout Design}
Data layout design is the core of FlashAccel co-design, because achieving high HBF bandwidth utilization depends not only on hardware capability, but also on whether data accesses can be arranged to match Flash parallelism. Efficient data layout enables data accesses in HBF to approach peak bandwidth. 

\subsection{Weights in HBF}

\begin{figure}[t]
    \centering
    \includegraphics[width=\linewidth]{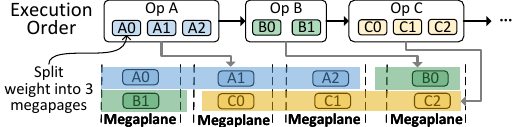}
    \caption{Data layout for weights in HBF.}
    \label{Fig::WeightLayout}
\end{figure}

FlashAccel places static weights across all megaplanes in execution order and aligns the layout with the hyperpage abstraction to exploit plane-level parallelism. Figure~\ref{Fig::WeightLayout} illustrates the weight layout strategy. For a given model, the computation graph of its inference step is fixed, which enables offline determination of the execution order. FlashAccel partitions each operator's static weights into megapage-sized chunks (e.g., A0, A1, and A2) and maps them across all megaplanes in round-robin order, packing them into consecutive hyperpages. After mapping the current operator, the system continues with subsequent operators according to the execution order, starting from the megaplane immediately following the last assigned megaplane of the current operator. As a result, the entire weight set is approximately balanced across megaplanes from a global perspective, and weight reads can be issued at hyperpage granularity to activate all planes concurrently.
Combined with the prefetch mechanism described in Section~\ref{Sec::Prefetch}, which loads these weights sequentially, this data layout enables high HBF bandwidth utilization.

\subsection{KV Cache in HBF} 
\label{Sec::KVCacheLayout}

\begin{figure}[t]
    \centering
    \includegraphics[width=.9\linewidth]{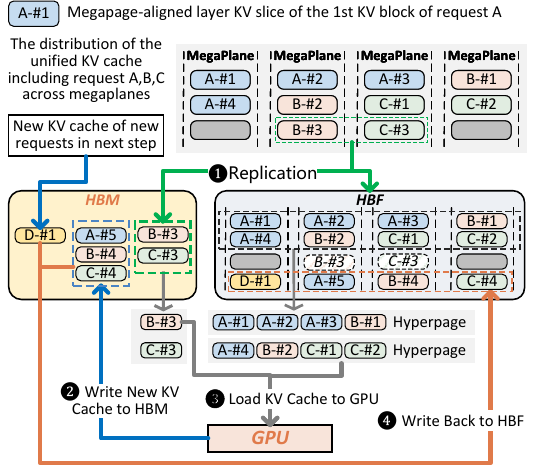}
    \caption{Data layout of KV cache in HBF and optimizations for achieving peak bandwidth.}
    \label{Fig::KVLayout}
\end{figure}

Compared with static weights, the layout of KV cache is more challenging because the active request set changes at every decode step. An HBF-friendly KV cache layout must meet two requirements. First, it should balance read and write load across planes so that no straggler plane limits the effective bandwidth. Second, it should align request-level KV cache eviction with block-granularity erase to avoid write amplification.
Following PagedAttention~\cite{kwon2023vllm}, the KV cache of a request is partitioned along the sequence dimension into logical \emph{KV blocks}. Each KV block contains the KV cache of $N$ consecutive tokens across all $L$ transformer layers and has a total size of $N\times L\times S$, where $S$ is the per-token KV cache size of one layer. Each logical KV block is further partitioned into $L$ \emph{layer KV slices}, each storing the KV data of the same $N$ tokens for one transformer layer, with a size of $N\times S$. FlashAccel aligns each layer KV slice to the megapage size, matching the layer-wise KV access pattern with the Flash read granularity.
Since the KV cache of all active requests is accessed together during inference, FlashAccel aggregates them into unified KV cache and addresses the straggler problem from a global perspective. FlashAccel adopts an isolated Flash block allocation policy, where each Flash block is allocated for the KV cache of only one request. Thus, evicting a request erases only its own Flash blocks, without migrating the data of other requests.

Figure~\ref{Fig::KVLayout} illustrates the layer-wise KV cache access flow during decode step. Before execution, FlashAccel performs selective replication to eliminate stragglers (Figure~\ref{Fig::KVLayout}~\ding{182}). Since each  layer KV slice in a KV block is aligned to megapage, all planes within a megaplane are accessed together and carry balanced loads, leaving the imbalance only among different megaplanes. Thus, FlashAccel measures the load distribution of the unified KV cache across megaplanes and copies excess megapages from megaplanes that hold more megapages than the average to HBM for each transformer layer. The original data remains in HBF, while accesses to these megapages are redirected to the copies in HBM. After replication, no megaplane's read load exceeds the target average load. This allows a larger fraction of HBF accesses to be issued at hyperpage granularity, reducing bandwidth loss caused by straggler planes. The replication overhead is small because the balanced layout is reused across decode steps, and each step only needs to handle the imbalance introduced by the changes in the active request set.

When executing a transformer layer, newly generated KV cache is first buffered in HBM (Figure~\ref{Fig::KVLayout}~\ding{183}). The attention kernel reads the KV cache in HBF at hyperpage granularity, while replicated and newly generated KV cache is accessed directly from HBM (Figure~\ref{Fig::KVLayout}~\ding{184}). One hyperpage access fetches multiple KV slices from different requests. Leveraging FlashAttention~\cite{dao2022flashattention}, FlashAccel computes partial attention results for individual KV slices and then reduces them by request to obtain the final attention output.
At the end of each step, FlashAccel writes eligible KV cache buffered in HBM back to HBF, including newly generated KV cache and KV cache of newly admitted requests in the next step (Figure~\ref{Fig::KVLayout}~\ding{185}).
However, directly writing KV cache from HBM to HBF would create an conflict between hyperpage granularity write and block isolation allocation. Once a request is allocated a megablock on a megaplane, its subsequent data must continue to be written to the same megaplane. Such request-megaplane binding prevents arbitrary redistribution of writes and makes writing HBF at hyperpage granularity difficult. Reserving one megablock per request on every megaplane would remove this constraint, but would cause severe fragmentation.

FlashAccel resolves this conflict with a megablock-level writeback policy. A KV block buffered in HBM becomes eligible for writeback once all $N$ token slots are filled, and eligible blocks from the same request continue to accumulate in HBM. Once they fill one or more megablocks, FlashAccel marks the request as writeback-ready and records the number of megablocks available for writeback. If the HBF contains $K$ megaplanes, FlashAccel initiates a writeback once $K$ megablocks are ready and writes $K$ megablocks to HBF in parallel. Specifically, each megaplane allocates one megablock, which is assigned to a single request and filled by its KV cache in one shot. Since every megaplane has exactly one megablock to write, the writeback can be issued at hyperpage granularity. This policy preserves isolated Flash block allocation while enabling all megaplanes to write simultaneously at hyperpage granularity. KV cache that has not yet satisfied the writeback condition remains buffered in HBM.

\begin{figure}[t]
    \centering
    \includegraphics[width=\linewidth]{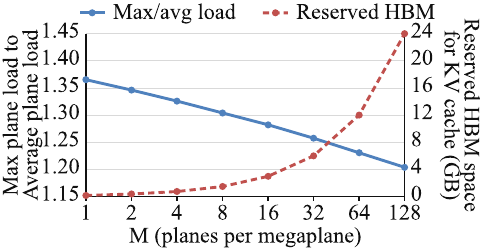}
    \caption{Tradeoff between HBF load balance and HBM KV cache reservation requirement under different megaplane granularities for Qwen3-235B with batch size 256 and sequence length 4K}
    \label{Fig::MegaplaneSize}
\end{figure}

Within the data layout, the megaplane granularity is a key parameter. A megaplane aggregates $M$ planes and exposes the aggregated megapage and megablock whose sizes grow with $M$. Increasing $M$ brings two benefits. First, more planes share one set of command control circuits, reducing the area overheads. 
Second, because load is inherently balanced within a megaplane, imbalance exists only across megaplanes.
Under the balls-into-bins model, larger megaplanes reduce the relative gap between the maximum and average megaplane load, which mitigates stragglers and improves bandwidth utilization. Considering an extreme case where a single megaplane contains all planes, no straggler plane exists.

However, a larger $M$ also increases HBM buffering pressure. The KV cache of a request can be written back only after its KV blocks are full and enough data has accumulated to fill a megablock. A larger $M$ brings a larger megablock size. It also increases the number $N$ of tokens in KV block, since the per-layer KV slice of a KV block is aligned with the larger megapage. Consequently, each request must buffer more KV cache in HBM before writeback, which is unfriendly to the HBM-constrained CLI architecture. Figure~\ref{Fig::MegaplaneSize} quantifies this tradeoff. We therefore set \(M=32\) planes per megaplane, which fundamentally alleviates the straggler load without imposing substantial buffering pressure on HBM.

\section{System Design}

The above data layout and access optimizations require system support at two levels. First, the system has to coordinate the accesses to HBM, HBF, and SRAM during inference. Second, it must organize persistent data stored in HBF. Accordingly, FlashAccel provides a programming model for heterogeneous resource coordination and an HBF-aware storage management layer for persistent data organization. 
The programming model exposes a unified virtual address space to programs. And the storage management layer maintains the metadata needed to locate persistent objects in HBF.

\begin{table*}[t]
\centering
\caption{Interface Overview}
\label{Tab::Interface}

\begin{tabular}{llll}
\hline
Interface    & Parameters                     & Return Value               & Description                                              \\ \hline
NandMmap     & objectID                       & (vaddr, size)              & Map an object in HBF into the virtual address (VA) space \\
GroupCreate  & objectIDs                      & groupID                    & Create a group object over multiple objects              \\
GroupAdd     & groupID, objectIDs             & groupID                    & Add objects to an existing group                         \\
GroupRemove  & groupID, objectIDs             & groupID                    & Remove objects from a group                              \\
GroupMmap    & groupID                        & (vaddr, size, readManifest) & Map a group object in HBF into the VA space              \\
GroupWrite   & groupID, writeManifest & -                          & Write new data into a group object                       \\
SramPrefetch & vaddr, prefetchSize            & vaddrSram                  & Prefetch data from Flash to SRAM via the mapped VA       \\
SramRelease  & vaddrSram                      & -                          & Release the SRAM space allocated by SramPrefetch         \\ \hline
\end{tabular}
\end{table*}

\subsection{HBF-aware Storage Management}


FlashAccel employs a lightweight HBF-aware storage management layer to manage the data resident in HBF.
This layer organizes model weights and KV cache as persistent objects, each identified by a unique object ID. In particular, the KV cache of each request is represented as a separate object. Each object is partitioned into fixed-size chunks at the granularity of megapages, and the storage management layer maintains per-object indexing metadata that records the physical address of every chunk. 

In conventional Flash management stacks, resolving an access to any part of an object requires two levels of address translation.
The program first obtains the logical block address (LBA) through the software-level filesystem.
Then, the LBA is translated by the Flash translation layer (FTL) in hardware (e.g., the SSD controller) into a physical Flash address.
Since Flash does not support in-place updates, the two-level indirection is necessary.
When data is modified, the new data must be written to a different physical address, and the FTL is responsible for tracking these remapped addresses.
However, the FTL is costly for HBF, as it demands a GB-scale logical-to-physical mapping table~\cite{gupta2009dftl} stored in low-latency SRAM or DRAM, and incurs extra address translation overhead~\cite{agrawal2008design}.

Our key observation is that model weights are immutable and KV cache follows append-only write patterns.
Once written, the physical address of each chunk remains unchanged.
FlashAccel therefore eliminates the FTL and records the physical address of each chunk directly in the object's indexing metadata, allowing the GPU to issue physical addresses to HBF without two-level address translation.
To persist the indexing metadata, FlashAccel adopts a log-based scheme~\cite{lee2015f2fs}.
Metadata updates are appended sequentially to dedicated log blocks in Flash rather than updated in place.
During system initialization, the storage layer replays the log to reconstruct the in-memory index of all objects, while periodic checkpointing compacts the log.

The management layer maintains the allocation state of all megablocks in HBF, tracking whether each megablock is free and, if not, the object that owns it. When allocating a megablock to an object, it follows the block isolation policy described in Section~\ref{Sec::KVCacheLayout}. Model weights and the KV cache of each request are placed in dedicated megablocks that are not shared across objects. Each megablock is exclusively assigned to a single object. As a result, reclaiming an object does not require relocating valid data belonging to other objects, eliminating write amplification and improving Flash endurance.

\subsection{Programming Model}

FlashAccel exposes HBF to programs through an interface set and a runtime. Table~\ref{Tab::Interface} summarizes the interfaces for accessing HBF resources. After the program invokes these interfaces, the runtime maintains GPU page table mappings from virtual addresses to the physical addresses of HBM, HBF, or SRAM. This design exposes heterogeneous memory through a unified virtual address space.
The programming model provides different interfaces for accessing weights and KV cache stored in HBF, accounting for distinct access characteristics.
To address the HBF access latency challenge, it also provides interfaces that leverage SRAM to hide data access latency.

\subsubsection{Weight Access Interface}

To efficiently manage read-only weight objects, FlashAccel provides a {\tt NandMmap} primitive that directly maps weight data stored in HBF into the virtual address (VA) range. Given a target weight object ID, {\tt NandMmap} retrieves the physical addresses of its constituent chunks via the object's indexing metadata, and updates the GPU page table to map the corresponding virtual addresses to these physical addresses. Subsequently, the program can seamlessly access the mapped weights via standard load operations on the returned virtual address. Since weights are static during execution, these mappings can be established once in model initialization and reused throughout inference.

\begin{figure}[t]
    \centering
    \includegraphics[width=\linewidth]{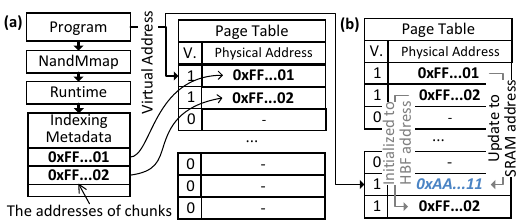}
    \caption{Page table mapping (a) after {\tt NandMmap}, and (b) further after {\tt SramPrefetch}.}
    \label{Fig::Nandmmap}
\end{figure}

\subsubsection{KV Cache Access Interface}

\begin{figure}[t]
    \centering
    \includegraphics[width=\linewidth]{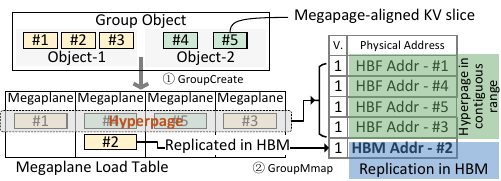}
    \caption{Group object construction and mapping for grouped KV cache access.}
    \label{Fig::KVAccess}
\end{figure}

FlashAccel manages each request’s KV cache as an independent object. However, during each inference step, the system must jointly read the KV cache of all active requests. To support this access pattern, we introduce a \textit{group object} abstraction. 
{\tt GroupCreate} logically combines the KV cache objects of active requests in the same inference step into a group object. As shown in Figure~\ref{Fig::KVAccess}, upon creation, the runtime traverses the metadata of all member objects and constructs a group-level \textit{megaplane load table} for each transformer layer. For each transformer layer and each megaplane, this table records the number and the physical addresses of its megapages containing data from the group object.
The membership of a group object changes as requests arrive and complete. {\tt GroupAdd} adds the KV cache objects of newly admitted requests to the group, whereas {\tt GroupRemove} removes the objects of completed request. Both operations incrementally update the megaplane load tables. Based on this abstraction, FlashAccel provides two data access interfaces, {\tt GroupMmap} and {\tt GroupWrite}, for grouped reads and append-only writes, respectively.

{\tt GroupMmap} maps a group object into a contiguous virtual address range and returns the base virtual address, mapped size, and a read manifest. Figure~\ref{Fig::KVAccess} illustrates the mapping process. When executing {\tt GroupMmap}, the runtime computes the average load across HBF megaplanes based on the megaplane load table and replicates excess megapages from overloaded megaplanes to HBM. It then updates the page table to map the virtual address range to the corresponding physical locations in HBF or HBM. During this process, the runtime iteratively selects one megapage from each megaplane based on the megaplane load table, packs them into a hyperpage, and maps these Flash megapages contiguously into the virtual address range. 
If certain megaplanes hold fewer megapages and cannot provide a megapage in subsequent rounds, the runtime instead selects a megapage replicated in HBM and maps the corresponding virtual address to its HBM location. This process repeats until all data has been mapped into the virtual address range. The runtime then returns a read manifest that maps each megapage-sized virtual address range to the object ID and chunk offset of its backing data. Before accessing a group object, the program should invoke {\tt GroupMmap} to build the mapping of the group object with its latest membership.

{\tt GroupWrite} flushes writeback-eligible KV cache data from HBM to HBF. It takes a \textit{write manifest} as input, which records the HBM locations of all KV cache chunks pending writeback and the object ID that each chunk belongs to. Following the design in Section~\ref{Sec::KVCacheLayout}, {\tt GroupWrite} first checks whether the pending data is ready for writeback and whether enough free megablocks are available. If free megablocks are insufficient, the runtime erases the KV cache object of the earliest requests to reclaim space. Once the write can proceed, the runtime allocates one megablock from each megaplane, and each allocated megablock is exclusively assigned to a single KV cache object. The runtime then writes data continuously at hyperpage granularity until all megablocks are full. After the write completes, the runtime updates the affected object metadata and the megaplane load table.

\begin{figure}[t]
    \centering
    \includegraphics[width=\linewidth]{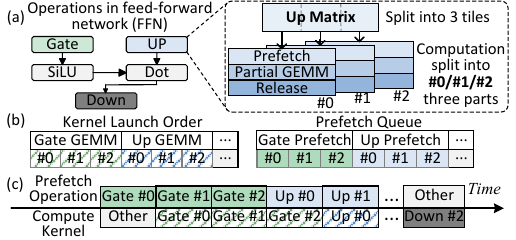}
    \caption{Proposed prefetch mechanism. (a) Modification to original compute kernels after introducing prefetch. (b) Kernel launch order and prefetch queue. (c) Overlap of prefetch and computation.}
    \label{Fig::Prefetch}
\end{figure}

\subsubsection{Data Prefetch}
\label{Sec::Prefetch}

FlashAccel incorporates a data prefetch mechanism into the programming model to leverage SRAM resources to hide the high read latency of Flash. 
Specifically, FlashAccel exposes the SRAM cache through programmer-controlled interfaces.
Programs explicitly invoke {\tt SramPrefetch}, which takes a virtual address referring to data in HBF and a prefetch size as input, and returns a virtual address referring to the prefetched data in SRAM for subsequent accesses.
FlashAccel also provides {\tt SramRelease} to release the allocated SRAM space after the last consumer finishes.
A kernel therefore follows a simple pattern: prefetch data to SRAM, compute on the returned address, and release the SRAM resources.
If the data larger than the SRAM capacity, program should split them into multiple smaller tiles that fit in SRAM and processes them sequentially. Figure~\ref{Fig::Prefetch}(a) illustrates how a large weight matrix is split into three tiles, each of which is prefetched into SRAM, multiplied with the activation, and released before processing the next tile.

To overlap prefetch with computation, the runtime employs prefetch-aware scheduling. Figure~\ref{Fig::Prefetch}(b) shows that {\tt SramPrefetch} operations are extracted from the compute graph, organized into a queue following the topological order of the computational graph, and executed asynchronously from the compute kernels. Whenever SRAM cache capacity becomes available, the runtime continuously dispatches prefetch requests from the head of the prefetch queue until the available SRAM space is exhausted. Multiple dispatched prefetch requests can execute concurrently. For each request, the runtime first establishes page table mappings for the returned virtual address range, with all entries initially pointing to the original data locations. 
It then attempts to allocate SRAM space from the SRAM cache within the circuit dies, resorting to the SRAM cache on the base dies if no space is available.
Once SRAM space is allocated, the runtime transfers the requested data from HBF to SRAM asynchronously from the compute kernels. A single prefetch request can access multiple megaplanes in HBF in parallel. Because the requested data is mapped to contiguous page table entries at hyperpage granularity, these accesses can be issued at the same granularity across megaplanes, allowing the prefetch operation to effectively exploit the bandwidth of HBF.
Prefetch does not stall dependent compute kernels. Before the data is prefetched, its page table entry retains the original HBF location. Once the prefetch completes, the runtime redirects the entry to the corresponding SRAM location. Compute kernels can always access the right data via the returned virtual address. Figure~\ref{Fig::Nandmmap}(b) illustrates the resulting page table mappings for a weight object during prefetch. The prefetch design allows Flash access latency to be overlapped with computation, as depicted in Figure~\ref{Fig::Prefetch}(c).

\section{Evaluation}
\subsection{Evaluation Setup}
\noindent\textbf{Workloads:} Table~\ref{Tab::Model} summarizes the four representative models used in our evaluation. These models cover popular attention mechanisms including group-query attention (GQA), multi-head latent attention (MLA)~\cite{liu2024deepseek-v2}, and feed-forward network (FFN) variants, including dense FFNs and mixture-of-experts (MoE) FFNs. We apply different parallelism strategies for these models including tensor parallelism (TP), data parallelism (DP) and expert parallelism (EP). FP16 is used for all models, except that the weights of DeepSeekV3~\cite{deepseekai2025deepseekv3technicalreport} are in FP8.
We set two sequence length configurations for different workloads. For long-context workloads, we derive the average sequence lengths of all tasks in the LongProc benchmark~\cite{ye2025longproc}, setting the input and output sequence lengths to 8.11K and 2.53K, respectively. The second configuration targets agentic workloads with multi-turn interactions. Based on statistics reported in \cite{wang2025agenttaxo}, we set the average input and output sequence lengths to 15K and 6K, respectively.

\begin{table}[t]
\centering
\caption{Model configuration}
\label{Tab::Model}
\fontsize{9pt}{11pt}\selectfont
\begin{tabular}{@{}ccc@{}}
\toprule
Models & Attention & FFN \\ \midrule
Qwen3-235B~\cite{yang2025qwen3} & GQA / DP @ 188KB/token & MoE / EP \\
Qwen3-Coder-480B~\cite{yang2025qwen3} & GQA / DP @ 248KB/token & MoE / EP \\
LLaMA3.1-405B~\cite{grattafiori2024llama} & GQA / TP @ 502KB/token & Dense / TP \\
DeepSeekV3-671B~\cite{deepseekai2025deepseekv3technicalreport} & MLA / DP  @ 70KB/token & MoE / EP \\ \bottomrule
\end{tabular}
\end{table}

\begin{figure*}[t]
    \centering
    \includegraphics[width=.95\textwidth]{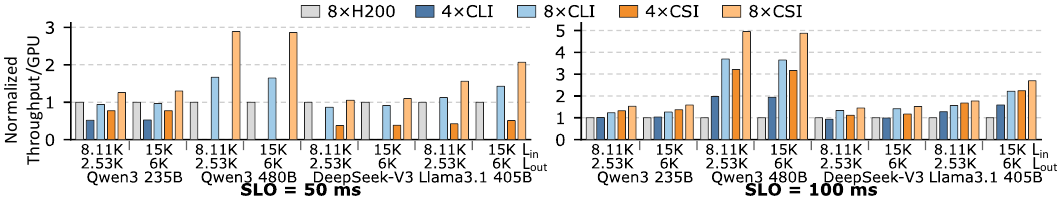}
    \caption{Throughput per GPU comparison among the baseline, CLI, and CSI under different models, sequence lengths, and latency SLOs .}
    \label{Fig::BigEval}
\end{figure*}

\noindent\textbf{Baseline and Configuration: }
The baseline is a DGX-H200 node~\cite{nvidia_dgx_h200_datasheet}. A DGX-H200 node contains 8 H200 GPUs within the same NVLink domain, each with 900 GB/s NVLink bandwidth.
Each H200 GPU is equipped with 6 HBM3e stacks~\cite{micron_hbm3e_ai}, delivering a total of 141GB DRAM capacity and 4.8 TB/s memory bandwidth. 
This baseline configuration is denoted as 8$\times$H200.

We model an HBF stack designed to deliver HBM3e-level bandwidth while providing 8$\times$ higher capacity. Table~\ref{Tab::Config} summarizes its configuration.
Following Lincoln~\cite{sun2025lincoln}, we design the Flash die based on XL-Flash~\cite{kouchi2020128gb}. Each Flash die contains 96 planes, and each plane has a capacity of 256MB. We derive area parameters from the die-shot in \cite{kouchi2020128gb}. For a conservative evaluation, $t_R$ and $t_{Prog}$ are kept the same as in \cite{kouchi2020128gb} even though the capacity of each plane is reduced by 4$\times$.
We integrate HBF into the GPU and evaluate two FlashAccel configurations: CSI and CLI.
CSI places 6 HBF stacks alongside the 6 HBM stacks in H200.
CLI replaces 5 of 6 HBM stacks with HBF and retains 1 HBM stack.
Since HBF offers higher capacity, fewer GPUs are needed to store model weights. Therefore, beyond 8$\times$CSI and 8$\times$CLI configurations, we also evaluate 4$\times$CSI and 4$\times$CLI.

\begin{table}[t]
\centering
\caption{HBF configuration.}
\label{Tab::Config}
\fontsize{9pt}{10.8pt}\selectfont
\begin{tabular}{>{\centering\arraybackslash}m{1cm}|>{\centering\arraybackslash}m{6cm}}
\hline
\multirow[c]{4}{1cm}[-1.5ex]{\centering\shortstack[c]{Array\\Die}}
& \shortstack[c]{\\[0.1ex]SLC 96 wordline-layers,\, 4KB / page} \\ \cline{2-2}

& \shortstack[c]{\\[0.1ex]192 pages/block , 342 blocks/plane  \\
96 planes (3 megaplanes), 192 Gb / die} \\ \cline{2-2}

& \shortstack[c]{\\[0.1ex]Plane area= 1.43 mm$^2$, TSV area = 12 mm$^2$,\\total area = 149 mm$^2$} \\ \cline{2-2}

& \shortstack[c]{\\[0.1ex]$t_R = 4\,\mu s$, $t_{Prog} =75\,\mu s$  } \\ \cline{2-2}

\hline

\multirow[c]{2}{1cm}[-2.2ex]{\centering\shortstack[c]{Circuit\\Die}}
& \shortstack[c]{\\[0.1ex]Plane peripheral circuit area = 1.21 mm$^2$,\\ 32KB SRAM per plane ( 0.050mm$^2$)} \\ \cline{2-2}

& \shortstack[c]{\\[0.1ex] Total area = 143 mm$^2$ (including TSV 12 mm$^2$ \\ and Other Peripheral area 10.17 mm$^2$) } \\ \hline

\shortstack[c]{\\[0.1ex]Base\\Die}
& \shortstack[c]{\\[0.2ex]SRAM capacity = 8MB } \\ \hline

\multirow[c]{2}{1cm}[-0.8ex]{\centering\shortstack[c]{HBF\\stack}}
& \shortstack[c]{\\[0.1ex]8 $\times$ Flash Die + 1 $\times$ Base Die @ 768GB/s} \\ \cline{2-2}

& \shortstack[c]{\\[0.1ex]Total Flash/SRAM capacity = 192GB / 32MB} \\ \hline
\end{tabular}
\end{table}

\noindent\textbf{Simulation:}
We build an event-driven simulator on top of LLMCompass~\cite{zhang2024llmcompass} to model the latency of LLM inference on both FlashAccel and conventional GPUs. LLMCompass searches for tiling strategies to find efficient GEMM mappings on GPU hardware and uses ScaleSim~\cite{samajdar2018scale} to estimate latency. We extend it with a NAND simulator that models megapage access latency at megaplane granularity using the plane-level timing parameters. We modify the memory model of LLMCompass to simulate the access of data residing in Flash and SRAM cache by providing the corresponding latencies.

\subsection{Evaluation Results}

Figure~\ref{Fig::BigEval} shows the throughput per GPU in the decode phase across five hardware configurations under varying models and sequence lengths. We use two decode latency service-level objectives (SLOs)~\cite{zhong2024distserve}, 50ms ($\sim$20 tokens/s/user) and 100ms ($\sim$10 tokens/s/user). 
FlashAccel alleviates the capacity bottleneck of HBM on batching, shifting the limiting factor for throughput from memory capacity to latency budget. Since larger batch sizes increase the arithmetic intensity of memory-bound FFN operations, this shift directly improves decode throughput. For each configuration, we therefore use the largest batch size that satisfies both the SLO and memory capacity constraints, and report throughput per GPU normalized to the baseline. For example, for LLaMA3.1-405B with 15K/6K context length, each request requires 10GB of KV cache, limiting the maximum batch size of 8$\times$H200 to 30, whereas under a 50ms SLO, 8$\times$CSI can support a theoretical batch size of 110 after accounting for the memory footprint of model weights.

Across all models, sequence lengths, and SLOs, 8$\times$CSI consistently outperforms the baseline, delivering a 2.10$\times$ average throughput gain by supporting substantially larger batch sizes.
8$\times$CLI shows lower throughput than 8$\times$CSI, as the CLI architecture has only 5 HBF stacks, resulting in 16.7\% lower bandwidth compared to CSI. Under the strict 50ms SLO, this bandwidth limitation causes it to slightly underperform the baseline on DeepSeek-V3 model and Qwen3-235B model. Under a 100ms SLO, the relaxed latency budget allows both CSI and CLI to load more KV cache per step and support larger batch sizes. As a result, 8$\times$CSI and 8$\times$CLI achieve 2.49$\times$ and 2.00$\times$ higher throughput per GPU than 8$\times$H200, respectively.

We further investigate configurations with fewer GPUs. For HBM-based GPUs, a 4$\times$H200 setup cannot fit the weights of most evaluated models and is not evaluated. Both 4$\times$CLI and 4$\times$CSI suffer from reduced aggregate bandwidth. For LLaMA3.1-405B and Qwen3-Coder-480B, loading weights alone exceeds 50ms, leaving insufficient time for KV cache loading and computation under a 50ms SLO. Thus, these models cannot be served under the evaluated configuration. For other models, the limited bandwidth and strict SLO restrict batch size, resulting in lower throughput than the 8$\times$H200 baseline.
When the SLO is relaxed to 100ms, more KV cache can be loaded per step, enabling larger batch sizes. Despite using fewer GPUs, FlashAccel still outperforms the baseline on LLaMA-405B, Qwen3-235B, and Qwen3-480B in throughput per GPU.

\begin{figure}[t]
    \centering
    \includegraphics[width=.9\linewidth]{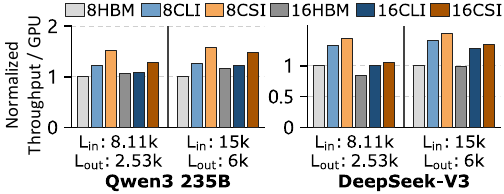}
    \caption{Throughput per GPU comparison after extending to 16 GPUs.}
    \label{Fig::Extend}
\end{figure}

Figure~\ref{Fig::Extend} shows throughput per GPU when scaling to 16 GPUs. For HBM-based GPUs, the additional memory capacity allows larger batch sizes. However, scaling to 16 GPUs moves communication from NVLink to RDMA~\cite{nvidia_dgx_h200_datasheet}, reducing bandwidth by about 9$\times$ and making all-to-all communication in EP much more expensive. As a result, the baseline shows little throughput improvement, and in some cases even degrades.
Although 16$\times$CSI and 16$\times$CLI still achieve higher throughput per GPU than 16$\times$H200, their performance is also limited by communication. Compared with 8$\times$CSI and 8$\times$CLI, both show lower throughput per GPU.
We also find that, for both CSI and CLI, an 8$\times$ HBF-based GPU system delivers higher throughput per GPU than a 16$\times$ HBM-based GPU system. Because each HBF-based GPU provides more capacity, a smaller system within a high interconnect bandwidth domain can sustain the workload, improving both cost efficiency and throughput per GPU.

\subsection{Ablation Study}

\begin{figure}[t]
    \centering
    \includegraphics[width=0.9\linewidth]{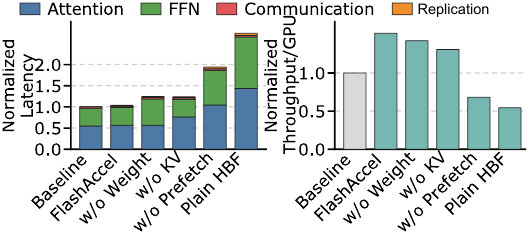}
    \caption{(a) Latency breakdown of different configurations. (b) Throughput comparison among different configurations.}
    \label{Fig::Ablation}
\end{figure}

We ablate each optimization in FlashAccel. Figure~\ref{Fig::Ablation} presents the latency breakdown of 6 accelerator configurations executing a Qwen3-235B model with a batch size of 256. Plain HBF denotes an HBF-based GPU without any optimizations. We further construct three ablated variants by disabling an optimization from the full CSI FlashAccel design at a time, namely prefetch, weight layout, and KV cache layout. With all optimizations enabled, FlashAccel matches HBM-based latency within $\sim$4\% at the same batch size.
Figure~\ref{Fig::Ablation} reports throughput under a 100ms SLO. Disabling prefetch, weight layout, and KV cache layout degrades throughput by 55\%, 7\%, and 15\%, respectively, relative to FlashAccel. Plain HBF suffers a 65\% throughput loss and falls below the HBM-based GPU. Overall, these results show that all three optimizations are necessary to close the performance gap of HBF.

\subsection{Write Issues}
Flash suffers from write limitations in both bandwidth and endurance. Table~\ref{Tab::Config} shows that the write latency of program operation ($t_{Prog}$) is significantly higher than the latency of read ($t_R$), which limits HBF write bandwidth. While the read bandwidth reaches 4.6 TB/s for CSI, the peak write bandwidth is only 245.8 GB/s. The endurance is another concern. Prior works~\cite{luo2015warm,cai2012flash} demonstrate that reducing retention time from 3 years to 3 days can extend P/E cycles by up to 50$\times$. 
Conservatively, we assume that reducing the retention time can extend the endurance by up to 10$\times$, increasing the P/E cycles from 100K to 1M. Even so, it still cannot match the nearly unlimited write endurance of DRAM. 
However, the wide adoption of new attention mechanisms, like GQA and MLA, reduces write volume compared to multi-head attention (MHA), making it possible to store KV cache in Flash. 
Across all evaluated configurations, each HBF-based GPU generates and writes at most 276MB of new KV cache per second in the decode node (8$\times$CSI running Qwen3-480B under a 100ms SLO, 1138 tokens/s). In addition, based on the input/output length ratio (i.e., $\sim$2.58:1) in agent workloads~\cite{wang2025agenttaxo}, each HBF-based GPU writes another 712MB of KV cache per second from the prefill node. In total, each HBF-based GPU writes 988MB of KV cache per second, taking only 3.9ms and incurring an additional overhead of 0.4\%.
Over a 5-year deployment period, this amounts to 148,570TB of total KV cache writes. With 1152GB capacity and 1M P/E cycles, CSI provides 1,125,000 TBW, sufficient for KV cache storage.

\subsection{Prefill and KV Cache Reuse}

\begin{figure}[t]
    \centering
    \includegraphics[width=.9\linewidth]{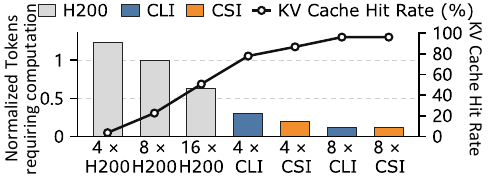}
    \caption{KV cache hit rate and normalized tokens requiring computation (normalized to 8$\times$H200).}
    \label{Fig::Hit}
\end{figure}

During the prefill phase, each request usually contains enough tokens to achieve high compute intensity, even without constructing a large batch size. In this case, both HBM-based GPUs and HBF-based GPUs can process the prefill requests efficiently.
However, in multi-turn interactions, the GPU for prefill needs to store the KV cache of previous turns for each session for future turns in the same session to reuse it, avoiding the recomputation for these cached tokens. 
Larger capacity allows the system to reduce KV cache eviction.
Figure~\ref{Fig::Hit} reports the KV cache hit rate measured using multi-turn interaction traces~\cite{badlogic_pi_mono} on different numbers of HBM-based GPUs and HBF-based GPUs. Due to the large memory capacity, both 8$\times$CLI and 8$\times$CSI can store the KV cache for all sessions and achieve the ideal hit rate. In comparison, even when we scale the HBM-based GPU system to 16 GPUs, its hit rate is still 50\% lower than that of the HBF-based GPU system. 
Figure~\ref{Fig::Hit} also shows that the higher hit rates of 8$\times$CSI and 8$\times$CLI reduce the number of tokens that require computation by up to 89\% compared with  8$\times$H200, thereby providing additional speedup.

\subsection{Energy and Area Analysis}
\begin{figure}[t]
    \centering
    \includegraphics[width=.95\linewidth]{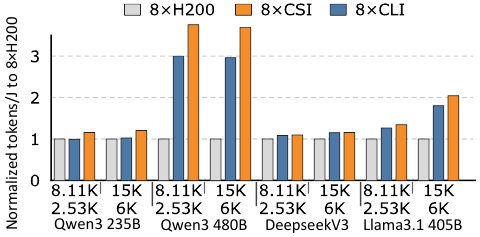}
    \caption{Energy efficiency comparison.}
    \label{Fig::Power}
\end{figure}

Flash memory suffers from high read energy consumption. While hybrid bonding can reduce read power, measurements from a real hybrid bonding Flash prototype still show a read energy of 8 pJ/bit~\cite{yanagidaira20251tb}, which is 2.68$\times$ higher than that of HBM3e (2.99 pJ/bit~\cite{antepara2025benchmark}). 
After integrating HBF, the overall TDP of CSI and CLI architectures increases 1.31$\times$ and 1.23$\times$, respectively. 
Nevertheless, thanks to the throughput advantages of HBF, we observe improvements in tokens per Joule.
Figure~\ref{Fig::Power} illustrates the energy efficiency (tokens/J) of different accelerators under a 100ms SLO. On average, 8$\times$CSI and 8$\times$CLI deliver 1.93$\times$ and 1.66$\times$ higher energy efficiency than 8$\times$H200, respectively.

Our evaluation shows that HBF occupies an area of 149mm$^2$, which is 1.23$\times$ larger than that of standard HBM3e (121mm$^2$). This overhead stems from our conservative assumptions based on the design in \cite{kouchi2020128gb}, which uses only 96 wordline layers. 
Recent technologies~\cite{cho2025321} have achieved over 300 wordline layers, significantly improving storage density and reducing area overhead for a given capacity. 
Using array dies with more wordline layers allows HBF to achieve area parity with the HBM stack. Even with only 96 wordline layers, our HBF design still achieves 6.50$\times$ higher density than HBM3e.

\section{Related Works}
\textbf{Flash Offloading Strategy:}
Prior works~\cite{sheng2023flexgen,hu2025tightllm,alizadeh2024llm} offload model weights or KV cache to SSDs to mitigate the DRAM capacity bottleneck, enabling LLM execution without out-of-memory failures. However, these designs are fundamentally constrained by the low SSD bandwidth, leading to extremely high decode latency, which limits their practicality in datacenter deployment. FlashAccel addresses this issue by increasing plane-level parallelism in Flash and employing dedicated layout optimizations to achieve high bandwidth.

\textbf{In-storage Processing Accelerators:}
These accelerators~\cite{lee2025aif,sun2025lincoln,yu2024cambricon} place compute units inside Flash to exploit its internal bandwidth for on-device inference. However, the limited silicon area inside Flash restricts computational capacity, resulting in low throughput. Moreover, they do not support storing KV cache in Flash, leaving capacity challenges unaddressed. FlashAccel instead integrates HBF with GPUs to provide high computational capacity, and enables KV cache storage via SLC Flash and system support for append-only KV cache writes.

\textbf{Processing-in-memory Accelerators:}
Some works~\cite{park2024attacc,fan2025sparseattentionremappingclustering} accelerate attention execution using DRAM-based processing-in-memory (PIM) which provides high internal bandwidth but limited computational capacity. They mainly target MHA that has low arithmetic intensity. Recent MLA and GQA impose higher computational demands that are difficult for PIM to sustain. In addition, PIM does not address the inherent low density of DRAM which leads to high cost. By contrast, FlashAccel improves density with Flash and, through GPU integration,  supports attention mechanisms requiring more computation.

\section{Conclusion}

This work points out the key obstacles that prevent HBF from translating its capacity advantages into serving throughput improvements, and addresses them through co-design across architecture, data layout, and system software.
FlashAccel introduces an SRAM cache at the architectural level and a prefetch interface at the system level to mitigate the high access latency of Flash. A specialized data layout fully leverages plane-level parallelism for weights and KV cache to improve bandwidth utilization. An HBF management layer and a programming model address the management and usage of heterogeneous memory resources.
Experimental results show that the CSI configuration of FlashAccel achieves 2.49$\times$ higher throughput than HBM-based GPU.

\section{Acknowledgment}

The authors acknowledge the use of GPT-5 and Kimi K2 for language polishing and grammar checking during manuscript preparation. GPT-5 and Opus-4 were used to assist with code review and bug fixing. All research ideas, technical designs, and code logic were independently developed by the authors. All AI-assisted text and code were carefully reviewed and revised by the authors. The authors take full responsibility for the content, methodology, and experimental results of this manuscript.

\bibliographystyle{IEEEtran}
\bibliography{refs}

\begin{IEEEbiography}
[{\includegraphics[width=1in,height=1.25in,clip,keepaspectratio]{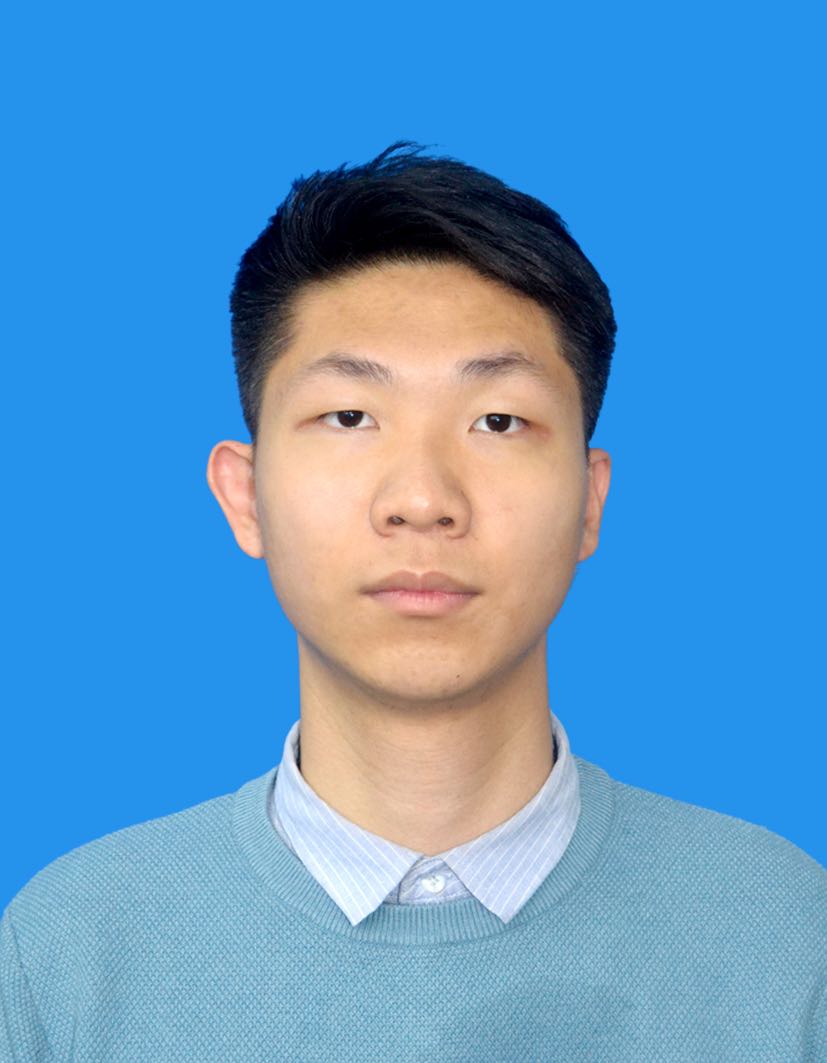}}]
{Xinyu Wang}
received the B.E. degree from the School of Computer Science and Technology,
Shandong University, Qingdao, China, in 2022. He is currently pursuing the
Ph.D. degree with the University of Chinese Academy of Sciences, Beijing,
China. His research interests include computer architecture,
processing-in-memory, and deep learning.
\end{IEEEbiography}

\begin{IEEEbiography}
[{\includegraphics[width=1in,height=1.25in,clip,keepaspectratio]{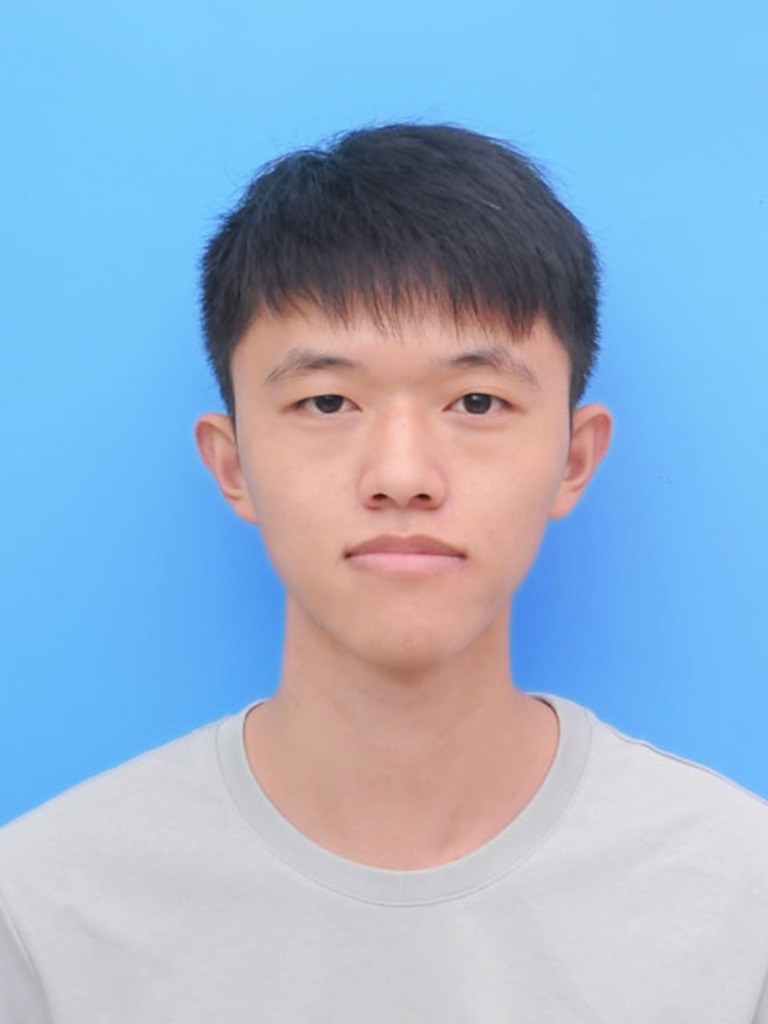}}]
{Yalong Xue}
 received the B.S. degree in physics from Shanghai Jiao Tong University, Shanghai, China, and is currently pursuing the Ph.D. degree in integrated circuit science and engineering with the University of Chinese Academy of Sciences, Beijing, China. His research interests include AI accelerator architecture, processing-in-memory (PIM), large language model inference acceleration and heterogeneous computing systems.
\end{IEEEbiography}

\begin{IEEEbiography}
[{\includegraphics[width=1in,height=1.25in,clip,keepaspectratio]{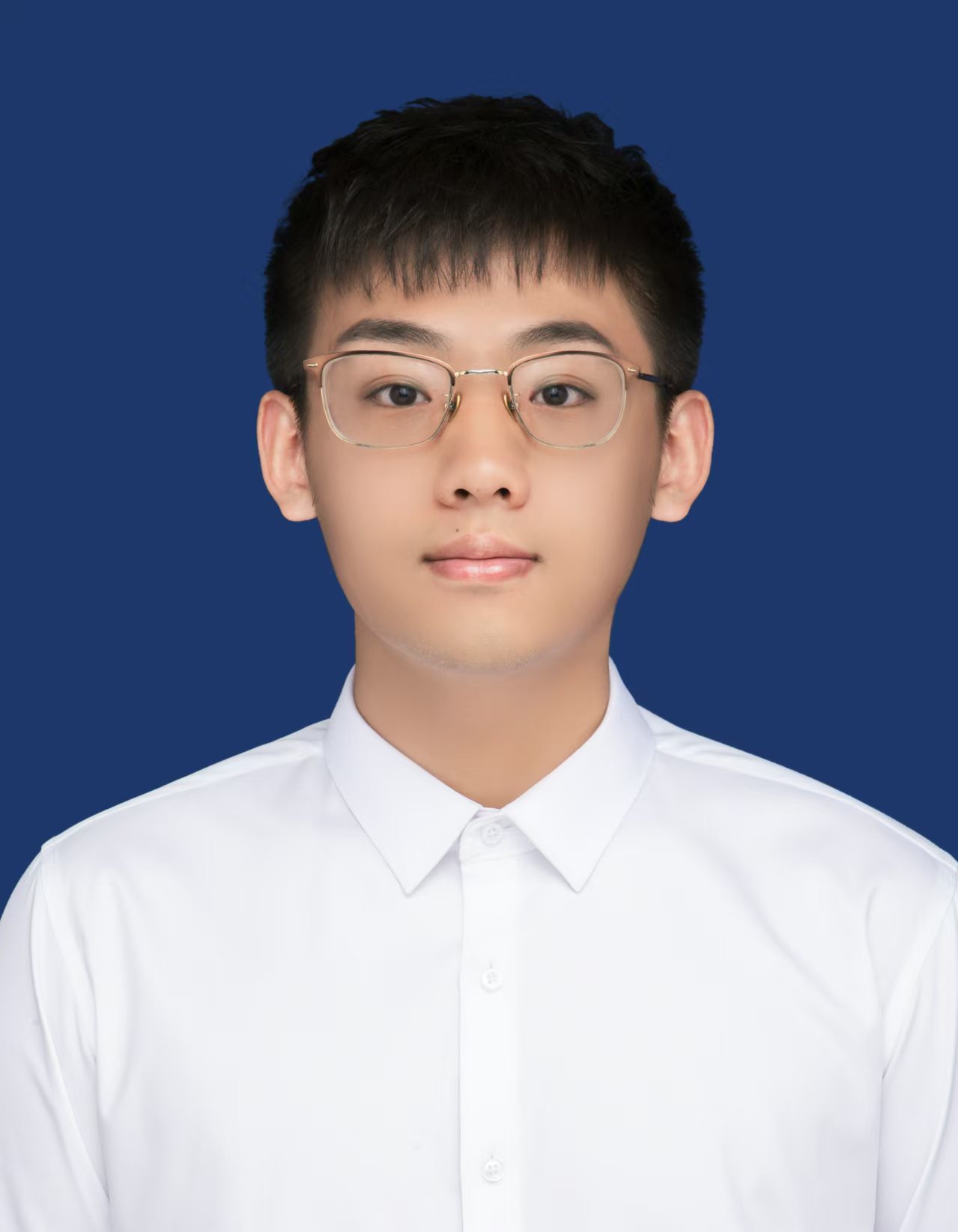}}]
{Xiaotian Sun}
received the B.S. degree in electronic engineering from Tsinghua University,
Beijing, China, in 2021. He is currently pursuing the Ph.D. degree with the
Institute of Computing Technology, Chinese Academy of Sciences, Beijing.
He is currently interested in accelerating deep neural networks on emerging
storage devices.
\end{IEEEbiography}

\begin{IEEEbiography}
[{\includegraphics[width=1in,height=1.25in,clip,keepaspectratio] {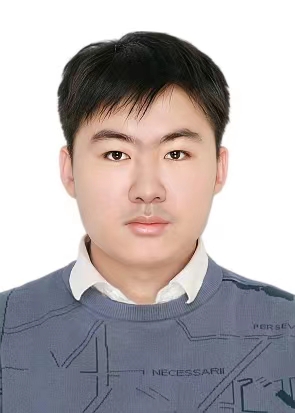}}]
{Xiaoyu Zhang}
received the Ph.D. degree from the University of Chinese Academy of Sciences, Beijing, China. Currently, he is an Assistant Professor with ICT, CAS. His research interests include processing-in-memory architecture design and electronic design automation.
\end{IEEEbiography}

\begin{IEEEbiography}
[{\includegraphics[width=1in,height=1.25in,clip,keepaspectratio] {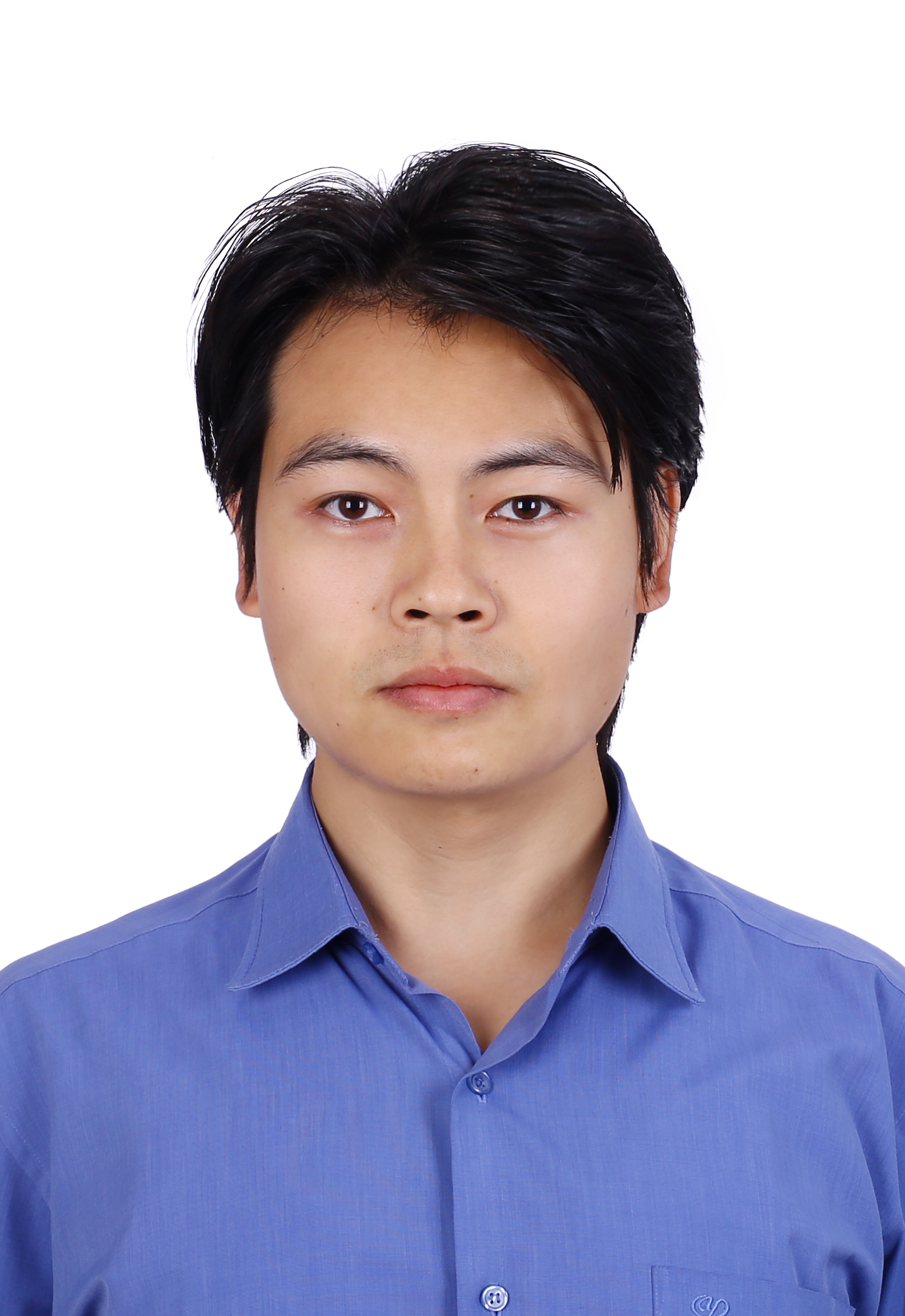}}]
{Xinjiang Zhang}
recived his BS in automation science and electrical engineering from Beihang University in 2015 and Ph.D. in physics from Beihang University in 2023. His research interests include quantum statistics, nano electronics, neuromorphic computing, and processing-in-memory, as well as computing systems based on heterogeneous processors. He is currently an Assistant Research Fellow at the Institute of Computing Technology, Chinese Academy of Sciences. 
\end{IEEEbiography}

\begin{IEEEbiography}
[{\includegraphics[width=1in,height=1.25in,clip,keepaspectratio]{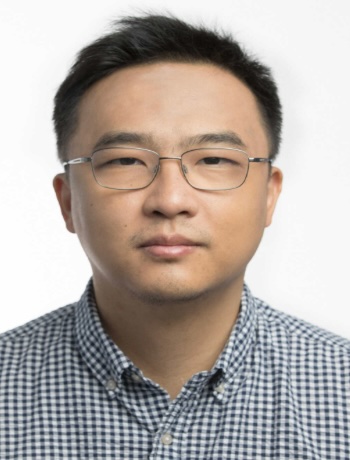}}]
{Chunmeng Dou}
received the B.S. degree in physics from Nanjing University, Nanjing, China,
in 2009, and the M.S. and Ph.D. degrees in electronics and applied physics
from the Tokyo Institute of Technology, Tokyo, Japan, in 2011 and 2014,
respectively. In 2018, he joined the Institute of Microelectronics, Chinese
Academy of Sciences, Beijing, China, where he is currently a Professor.
His research interests include emerging memory circuit design, design
technology co-optimization, and emerging computing paradigms.
\end{IEEEbiography}

\begin{IEEEbiography}
[{\includegraphics[width=1in,height=1.25in,clip,keepaspectratio]{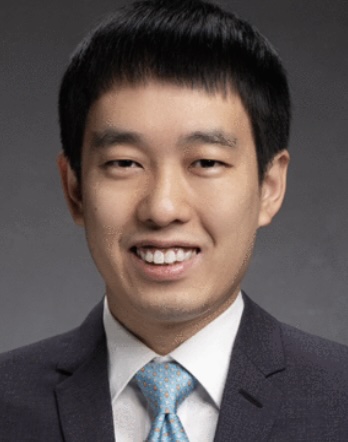}}]
{Xueqi Li}
received the Ph.D. degree from Institute of Computing Technology (ICT),
Chinese Academy of Sciences (CAS), under the supervision of Prof. Ninghui Sun.
From 2018 to 2020, he was also co-advised by Prof. Yuan Xie as a joint PhD
student with the SEAL Lab, UCSB. He is currently an Associate Professor at
the ICT, CAS. His research interests include Domain-specific PIM accelerators
and AI for Life Sciences.
\end{IEEEbiography}

\begin{IEEEbiography}
[{\includegraphics[width=1in,height=1.25in,clip,keepaspectratio]{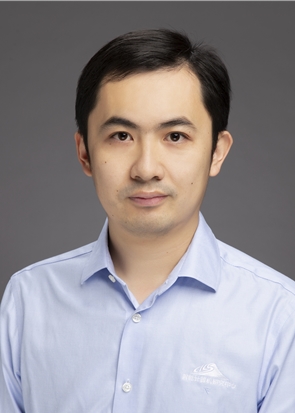}}]
{Xiaoming Chen}
(Member, IEEE) received the B.S. and Ph.D. degrees in electronic engineering
from Tsinghua University, Beijing, China, in 2009 and 2014, respectively.
He is currently a Professor with the Institute of Computing Technology,
Chinese Academy of Sciences, Beijing. His current research interest is focused
on design automation for integrated circuits and PIM architectures.
\end{IEEEbiography}

\end{document}